\documentclass[aps,superscriptaddress,nofootinbib,eqsecnum,prd,notitlepage,twocolumn]{revtex4-1} 

\pdfoutput=1

\usepackage{amsfonts}
\usepackage{amsmath}
\usepackage{amssymb}
\usepackage{graphicx,color}
\usepackage{float}
\usepackage{hyperref}
\usepackage{subfigure}
\usepackage{dcolumn}% Align table columns on decimal point
\usepackage{soul}
\usepackage{ulem}

\begin{document}

\title{Unified early and late Universe cosmology through
  dissipative effects in steep quintessential inflation potential models}

\author{Gustavo B. F. Lima}
%\email{}
\affiliation{Observat\'orio Nacional, 20921-400 Rio
de Janeiro, RJ, Brazil}

\author{Rudnei O. Ramos}
%\email{rudnei@uerj.br}
\affiliation{Departamento de Fisica Teorica, Universidade do Estado do Rio de Janeiro, 
20550-013 Rio de Janeiro, RJ, Brazil }

%%%%%%%%%%%%%%%%%%%%%%%%%%%%%%%%%%%%%%%%%%%%%%%%%%%%%%%%%%%%%%%%%%%%%%%%%%%
\begin{abstract}

By making use of a class of a steep exponential type of potentials,
which has been recently used to describe quintessential inflation, we
show how a unified picture for inflation, dark energy, and dark
matter can emerge entirely through dissipative effects.
Dissipation provides a way to extend the applicability of a larger
class of these potentials in the sense of leading to a consistent
early Universe inflationary picture and producing observables in
agreement with the Planck legacy data. Likewise, dissipative effects
lead to dark matter production with consistent abundances and, toward
the recent time of the  Universe, drives the potential energy of the scalar
quintessential field to dominate again, essentially mimicking a cosmological 
constant by today, with all cosmological parameters consistent with the 
observations. Both early and late Universes are connected and have no 
kination period in between.   

\end{abstract}
%%%%%%%%%%%%%%%%%%%%%%%%%%%%%%%%%%%%%%%%%%%%%%%%%%%%%%%%%%%%%%%%%%%%%%%%%%%

\maketitle

%%%%%%%%%%%%%%%%%%%%%%%%%%%%%%%%%%%%%%%%%%%%%%%%%%%%%%%%%%%%%%%%%%%%%%%%%%
\section{Introduction} 

By now, we have accumulated a large number of cosmological data coming
from different sources and at increasing precision, as the ones
obtained by the most recent cosmological probes,  including data from
the cosmic microwave background (CMB)
anisotropies~\cite{Aghanim:2018eyx}, galaxy clustering, e.g., baryon
acoustic oscillations
(BAO)~\cite{Anderson:2013zyy,Kazin:2014qga,Bautista:2017wwp},
gravitational lensing~\cite{Abbott:2017wau}, the large-scale structure
(LSS) of the Universe~\cite{Scolnic:2017caz} and
supernovae~\cite{Abbott:2018wog}.  Putting all together, the
observational data point to a Universe that fits  the
so-called concordance model of cosmology, or $\Lambda$CDM model, quite well. The
$\Lambda$CDM model is the simplest cosmological model fitting existing
data and describes our Universe in terms of a cosmological constant
$\Lambda$ modeling dark energy for the recent accelerating expansion
and cold dark matter (CDM),  as the dominant energy components.
{}Finally, the $\Lambda$CDM model is complemented by the idea of
inflation as the paradigm to the solution of the big bang cosmological
model for the early Universe.

Though inflation, dark energy and dark matter are considered as
separated entities in general in the literature, there are proposals
formulating different unified pictures for
them~\cite{Capozziello:2005tf,Liddle:2008bm,Henriques:2009hq,Bose:2009kc,DeSantiago:2011qb,Guendelman:2016kwj}.
This is motivated by trying to have a simple picture for these
different forms of energy that have prevailed in the early Universe
(inflation) and in the recent cosmological history (dark energy and
dark matter).  In fact, the $\Lambda$CDM model itself motivates these
studies, remembering that the model itself faces unsolved theoretical
issues, like the fine-tuning problem~\cite{Copeland:2006wr}, the
cosmic coincidence problem~\cite{Steinhardt:1999nw} and the
unexplained numerical value of the cosmological constant itself. 

In the present work, we propose a model implementing a typical
quintessential inflation behavior, i.e., through a scalar field model
that can play both the roles of the inflaton and the dark energy (in
the form of a dynamical scalar-quintessence field), in the early
and late cosmology histories of the Universe,
respectively. Quintessential inflation models are abundant in the
literature~\cite{Peebles:1998qn,Dimopoulos:2001ix,Rosenfeld:2005mt,Nojiri:2005pu,BuenoSanchez:2006fhh,Rosenfeld:2006hs,BasteroGil:2009eb,Ito:2011ae,Bamba:2012cp,Hossain:2014zma}.
One of the main novelties in the model to be presented here is the role that
dissipation will be making throughout the dynamics, from the very
early inflationary Universe until the recent dark energy accelerated
epoch. In addition, dissipation itself will serve us in producing the
appropriate abundance of dark matter, thus naturally and nicely
connecting the end of inflation with the big bang early radiation
dominated Universe, passing by the matter-dominated regime and ending
in the recent  (dark energy-dominated) acceleration regime. We will
realize this program in the context of the warm inflation
picture~\cite{Berera:1995ie}  (see also
Refs.~\cite{Berera:2008ar,BasteroGil:2009ec} for reviews), which also
serves to motivate how energy can be exchanged in between the dark
sector components (i.e., between the quintessential inflaton field and
the dark matter).

Recently, quintessential inflation models motivated from warm
inflation have been suggested~\cite{Dimopoulos:2019gpz,Rosa:2019jci}.
These models have a number of attractive features. {}For example, one
does not need to rely on gravitational particle production  in these
models as the leading mechanism for generating the radiation-dominated
regime after inflation\footnote{
Note, however, that there are other ways of reheating the Universe in
quintessential inflation and that do not rely on gravitational particle 
production, like instant preheating~\cite{Agarwal:2017wxo,Dimopoulos:2017tud},
curvaton reheating~\cite{BuenoSanchez:2007jxm,Matsuda:2007ax,Qiu:2016mrx}, 
Ricci reheating~\cite{Dimopoulos:2018wfg,Opferkuch:2019zbd} and 
reheating with a trap~\cite{Dimopoulos:2019ogl}.
}, like in the original quintessential inflation
model by Peebles and Vilenkin~\cite{Peebles:1998qn}. Besides, on a
more fundamental level, there has recently been increased interest in
how present field theory models describing either inflation, or dark
energy, or both, can be accommodated in a consistent ultraviolet
completion in quantum gravity/string theory. This gave origin to the
recent so-called  swampland conjectures (see, e.g.,
Ref.~\cite{Palti:2019pca} for a thorough recent review and
references therein), which puts in check the $\Lambda$CDM model, dark
energy models and inflation models alike. In this perspective, warm
inflation  has been discussed as a possible and natural way to evade
the issues brought about by these swampland
conjectures~\cite{Motaharfar:2018zyb,Das:2018rpg,Bastero-Gil:2019gao}.

While the authors in Refs.~\cite{Dimopoulos:2019gpz,Rosa:2019jci}
formulated the quintessential inflation model in the warm inflation
context with the Peebles and Vilenkin type of
potential~\cite{Peebles:1998qn}, here, instead, we will make use of a
class of steep exponential potentials for quintessential
inflation~\cite{Geng:2015fla,Geng:2017mic,Ahmad:2017itq,Shahalam:2017rit,Das:2019ixt}.
Contrary to the thawing type of models for quintessence, this is a
freezing type of model (for a distinction between thawing and freezing
models, see, e.g., Ref.~\cite{Caldwell:2005tm}). While thawing models
are very sensitive to the initial conditions, the freezing models are
essentially independent of the initial
conditions~\cite{Tsujikawa:2013fta}.  {}Freezing models can also
exhibit a tracker or a scaling behavior, depending whether they can
provide a late-time acceleration  or not, while this is not possible
in scaling models. The simple exponential potential, for example,
exhibits a tracking behavior and as such, it does not lead to a
correct present-time behavior in which the equation of state of dark
energy should be such that $\omega_\phi \simeq -1$ and it is then
considered in general only as a suitable quintessence model with
modifications of its potential~\cite{Copeland:2006wr,Chiba:2012cb}.
{}For the steeper potential of a generalized exponential form as
introduced in  Refs.~\cite{Geng:2015fla,Geng:2017mic,Ahmad:2017itq},
the energy density of the scalar field freezes in the past because of
the Hubble damping, and later on it starts evolving and  scales with the
background toward the recent epoch, after which it exits to the
background providing then the late-time acceleration recently. Besides
having all the desirable properties as a quintessence model, it also
produces a consistent early Universe inflationary cosmology in
agreement with the Planck data, as also shown in
Refs.~\cite{Geng:2015fla,Geng:2017mic,Ahmad:2017itq}. 

To connect with dark matter,  the authors in Ref.~\cite{Geng:2017mic}
made use of the so-called growing neutrino quintessence
model~\cite{Wetterich:2007kr,Amendola:2007yx}.  Because of the still
approximate scaling behavior of the model, by coupling the scalar
field non minimally to neutrinos makes  their mass to grow as the
scalar field grows and this helps to trigger the late-time transition
to the present accelerated expansion epoch.  Here, instead, we couple
the inflaton quintessential field to matter only through energy
exchange effects. This, as we are going to demonstrate, when done in
an appropriate way, this leads naturally to a matter-dominated regime and
at the same time  will make the scalar field energy density
dominate again at later time, closer to the recent epoch, triggering
the transition from  matter domination to the current accelerated
regime. This approach is very much similar to the recent studies on
dark matter-dark energy
interactions~\cite{Amendola:1999er,delCampo:2006vv,Amendola:2006qi,Chimento:2007yt,Rosenfeld:2007ri,delCampo:2008sr,delCampo:2008jx,Quartin:2008px,Chimento:2009hj,Duran:2010hi,Arevalo:2011hh,Zimdahl:2014jsa,
  Pu:2014goa,Abdalla:2014cla,Shahalam:2015sja,Santos:2017bqm,Yang:2018xlt,vonMarttens:2018iav,Benetti:2019lxu}.
This type of approach, in which  interactions in the dark
sector are assumed, has many appealing properties like, for example, providing a way
of explaining the coincidence problem or at least  alleviating it (for
a thorough description of the properties of these dark sector
interacting models, see, e.g., the review papers
Refs.~\cite{Bolotin:2013jpa,Wang:2016lxa}). Typically, the proposed
forms for the interaction terms involving the dark sector are in
general purely phenomenological, with many different forms treated in
the literature. Here, however, we will motivate the interactions fully
from the warm inflation picture. In warm inflation, different forms of
dissipation coefficients have been considered (for examples, see,
e.g., Refs.~\cite{BasteroGil:2010pb,BasteroGil:2012cm}). It happens
that depending on the form of the scalar potential and the dissipation
coefficients, different dynamical behaviors can emerge.  In the
present work, we will use for the inflationary part of the dynamics
the dissipation coefficient described in
Refs.~\cite{Berera:2008ar,BasteroGil:2012cm}, which combined with the
generalized exponential form for the potential, has the property of
naturally becoming negligible at the end of inflation and no further
significant radiation produced from the energy of the
inflaton/quintessence scalar field to light degrees of freedom being
generated later on. However, we can still have interactions between
the scalar field and the produced nonrelativistic matter
afterward. These interactions, in fully analogy with the behavior
expected from similar dissipation processes in the warm inflation
dynamics, can be chosen such as to initially promote the growth of the
nonrelativistic matter over radiation, creating a long enough matter-domination 
period after radiation domination. Then, it will
subsequently contribute to making the scalar field energy density to
dominate again in a mechanism similar to the one observed with other
approaches with interactions in the dark sector.

The next sections of this work are organized as follows. In
Sec.~\ref{sec2}, we will present the model that we study here and its
implementation in the warm inflation picture. In Sec.~\ref{sec3}, we
will study the early Universe inflationary dynamics of the model and
give its observational predictions, contrasting them with the Planck
legacy data. We will dedicate  Sec.~\ref{sec4} to the study of the
late Universe dynamics and also obtain the different predictions of
the model in this epoch. We will show that the model provides a fully
consistent picture in this epoch and also contrast the results, at the
background level, with those from the Planck data.  {}Finally, in
Sec.~\ref{sec5}, we will give our conclusions.    

%%%%%%%%%%%%%%%%%%%%%%%%%%%%%%%%%%%%%%%%%%%%%%%%%%%%%%%%%%%%%%
\section{Model}
\label{sec2}

In the warm inflation (WI) dynamics~\cite{Berera:1995ie} the inflaton,
described by a scalar field $\phi$, is coupled to radiation bath
degrees of freedom, with energy density $\rho_R$, through a
dissipation type of coefficient that is fully motivated and can be
derived from microscopic physics from quantum field
theory~\cite{Berera:2008ar}. The background dynamical evolution is
defined by the equations
\begin{eqnarray}
&& \ddot \phi + 3 (1+Q) H \dot \phi + V_{,\phi}=0,
\label{eqphi}
\\ && \dot \rho_R + 4 H \rho_R = 3H Q \dot \phi^2,
\label{eqrhoR}
\end{eqnarray}
where dots denote temporal derivatives, $H$ is the Hubble parameter,
\begin{equation}
H^2 \equiv \left(\frac{\dot a}{a}\right)^2 =\frac{1}{3 M_{\rm Pl}^2}
\rho,
\label{Hubble}
\end{equation}
with $\rho$ the total energy density, $a\equiv a(t)$  is the scale
factor, $M_{\rm Pl} \equiv 1/\sqrt{8 \pi G} \simeq 2.4 \times
10^{18}$GeV is the reduced Planck mass and $Q$ in Eqs.~(\ref{eqphi})
and (\ref{eqrhoR}) is the dissipation ratio in WI, defined as
\begin{equation}
Q= \frac{\Upsilon(T,\phi)}{3 H},
\label{Q}
\end{equation}
where $\Upsilon(T,\phi)$ is the dissipation coefficient in WI, which
can be a function of the temperature and/or the background inflaton
field, depending on the microscopic physics for WI. {}For instance, in
Refs.~\cite{Berera:2008ar,BasteroGil:2010pb,BasteroGil:2012cm,Bastero-Gil:2016qru,Bastero-Gil:2019gao}
there are many examples of models leading to different forms for
$\Upsilon(\phi, T)$.  {}For example, in the models of
Refs.~\cite{Berera:2008ar,BasteroGil:2010pb,BasteroGil:2012cm}, the
dissipation coefficient typically scales with the temperature $T$ of
the radiation bath and the background inflaton field as $\Upsilon
\propto T^3/\phi^2$. In the model studied in
Ref.~\cite{Bastero-Gil:2016qru} we find instead that $\Upsilon \propto
T$, while in the more recent construction done in
Ref.~\cite{Bastero-Gil:2019gao}, it is found that $\Upsilon \propto
M^2/T$, where $M$ is a mass scale in the model.  Motivated by these
functional forms found for the dissipation coefficient in WI, we can
make a generic parametrization for  the dissipation coefficient like
\begin{equation}
\Upsilon(T,\phi) = C T^c \phi^p M^{1-c-p},
\label{Upsilon}
\end{equation}
where $C$ is a dimensionless constant (that carries the details of the
microscopic model used to derive the dissipation coefficient, e.g.,
the different coupling constants of the model) and numerical powers
given by $c$ and $p$, which can be either positive or negative
numbers (note that the dimensionality of the dissipation coefficient
in Eq.~(\ref{Upsilon}) is preserved, $[\Upsilon] = [{\rm energy}]$).  

{}Furthermore, we recall that the slow-roll parameters in WI are
modified with respect to the ones in the cold inflation scenario to
\begin{eqnarray}
\epsilon_{WI} &=& \frac{\epsilon_V}{1+Q} ,
\label{eps}
\\ \eta_{WI} &=& \frac{\eta_V}{1+Q},
\label{eta}
\end{eqnarray} 
where
\begin{eqnarray}
\epsilon_V &=& \frac{M_{\rm Pl}^2}{2} \left( \frac{V_{,\phi}}{V}
\right)^2 ,
\label{epsV}
\\ \eta_V &=& M_{\rm Pl}^2 \frac{V_{,\phi\phi}}{V}.
\label{etaV}
\end{eqnarray} 
 
Given the dissipation coefficient expressed like Eq.~(\ref{Upsilon})
and using the slow-roll approximations for the equations
(\ref{eqphi}), (\ref{eqrhoR}) and (\ref{Hubble}), we can deduce, after
some algebra, that $Q$ and $T/H$ have evolution equations, expressed
in terms of the number of e-folds, $dN = H dt$, given, respectively, by
\begin{widetext}
\begin{eqnarray}
\frac{d \ln Q}{dN} &=& \frac{2\left[ (2 + c) \epsilon_{V} - c \,
    \eta_{V} -  2 p \kappa_{\rm V} \right]}{4 - c + (4 + c) Q},
\label{dQdN}
\\ \frac{d\ln (T/H)}{dN} &=& \frac{ \left[7 + c (Q-1) + 5 Q\right]
  \epsilon_{V} -  2 (1 + Q) \eta_{V} + (Q-1)p \kappa_{V} } {(1 + Q) [4
    - c + (4 + c) Q]},
\label{dTHdN}
\end{eqnarray}
\end{widetext}
where $\kappa_{V}$ is defined as
\begin{equation}
\kappa_{V} =  M_{\rm Pl}^2 \frac{V_{,\phi}}{\phi V}.
\label{kappa}
\end{equation}
The above equations determine the evolution of $Q$ and $T/H$ in the WI
dynamics.  Depending on the scalar field potential model and
dissipation coefficient, we can have different behaviors for these
quantities, for example, growing or decreasing with the number of
e-folds whether the right-hand-sides of the Eqs.~(\ref{dQdN}) and
(\ref{dTHdN}) are positive or negative, respectively. In particular,
the denominator of these equations can be shown to be always positive
and this is ensured when studying the dynamical stability in general
of the background equations in
WI~\cite{Moss:2008yb,delCampo:2010by,BasteroGil:2012zr}. These studies
have shown that WI has a stable dynamics provided that the power $c$
in the dissipation coefficient Eq.~(\ref{Upsilon}) satisfies $-4 < c <
4$, which is valid in the weak ($Q \ll 1$) and strong ($Q>1$)
dissipative regimes of WI.  The sign of the numerator in
Eqs.~(\ref{dQdN}) and (\ref{dTHdN}) can also be determined according
to the model in study. 

As already explained in the Introduction, in this work we will be
working with the scalar potential for quintessential inflation given
by the generalized exponential
form~\cite{Geng:2015fla,Geng:2017mic,Ahmad:2017itq}
\begin{equation}
V(\phi) = V_0 \exp\left[-\alpha (\phi/M_{\rm Pl})^n\right],
\label{pot}
\end{equation}
where $V_0$ is the normalization of the potential and $\alpha$ is a
dimensionless constant. In the conventions used in the present work,
$\alpha$ is considered as a positive number and  the time derivative
of the scalar field is also positive, $\dot \phi >0$. We are also
interested in potentials steeper than the simple exponential one,
hence, $n>1$.  When $\alpha (\phi/M_{\rm Pl})^n \ll 1$, the potential
is sufficiently flat to allow inflation, while as $\phi$ rolls down
the potential and $\alpha (\phi/M_{\rm Pl})^n$ becomes closer to 1,
the potential steepens, ending inflation. {}Finally, as $\phi$
continues to increase, for $\alpha (\phi/M_{\rm Pl})^n \gtrsim 1$, a
late-time scaling solution emerges. As discussed and explained by the
authors in Ref.~\cite{Geng:2017mic}, the scalar field $\phi$ can
exceed the Planck scale when $\alpha \ll 1$, yet with no related
ultraviolet (UV) issues.

Since the dissipation coefficient in WI is in general a function of
the temperature, it couples the scalar quintessential inflaton field
directly with the radiation.  Since from the observational point of
view there is no reason to have this coupling  present in the late
Universe, the dissipation coefficient (\ref{Upsilon}) must be relevant
only at the early times and be negligible afterward. This is
equivalent to requiring that $Q$ should be a decreasing function of time
(number of e-folds) by the epoch that the scalar field enters in the
scaling regime where  $\alpha (\phi/M_{\rm Pl})^n \gtrsim 1$.
Substituting Eq.~(\ref{pot}) in Eq.~(\ref{dQdN}) for instance, we find
\begin{eqnarray}
\frac{d \ln Q}{dN} &=& - \frac{n \alpha \left(\frac{\phi}{M_{\rm
      Pl}}\right)^{n-2} } {4 - c + (4 + c) Q} \nonumber \\ &\times&
\left[ -2 c (n-1)-4p +(c-2) n \alpha \left(\frac{\phi}{M_{\rm
      Pl}}\right)^n \right].  \nonumber \\
\label{dQdNexpn}
\end{eqnarray}
Thus, we find that in the scaling regime and  for $c>2$, the
right-hand-side of Eq.~(\ref{dQdNexpn}) is negative, hence, $Q$ is a
decreasing function with the number of e-folds. This is in particular
the case for the dissipation coefficient of
Refs.~\cite{Berera:2008ar,BasteroGil:2010pb,BasteroGil:2012cm}, i.e.,
when considering $c=3,\, p=-2$ in  Eq.~(\ref{Upsilon}). One should
also note that there are other ways of making $\Upsilon$ vanish
at later times, after the inflationary regime, due to the intrinsic
microscopic details of the WI construction, like in
Refs.~\cite{Bastero-Gil:2016qru,Bastero-Gil:2019gao}, in which after the
temperature decreases to a value below a characteristic scale of the model, the
dissipative effects can shut down. However, this is mostly dependent of
the microscopic physics involved and in the present work, we adopt a
more model independent approach, in which the coupling between the scalar
field and radiation can naturally become negligible (and effectively
inefficient) after inflation as a result of the background dynamics.  Hence,
from now on we will assume that the early Universe inflationary
dynamics is dominated by the dissipation coefficient with the cubic
dependence on the temperature,
\begin{equation}
\Upsilon_{\rm cubic} = C_{\rm cubic} \frac{T^3}{\phi^2},
\label{cubic}
\end{equation}
and the dissipation ratio that couples the scalar quintessential
inflaton field to radiation is then $Q \equiv \Upsilon_{\rm cubic}/(3
H)$.

In addition to Eqs.~(\ref{eqphi}) and (\ref{eqrhoR}) for the
background dynamics of $\phi$ and $\rho_R$, we should also complement
them with the evolution equation for the matter energy density,
$\rho_m$, so to be able to properly describe the late-time evolution
of the Universe.  The matter energy density can also be split into the
dark matter and baryon energy densities, $\rho_{DM}$ and $\rho_b$,
respectively, such that $\rho_m \equiv \rho_{DM} + \rho_b$.  In the
present work we will not make the distinction between the background
evolution for each of these components separately and we will treat
only the evolution of the total matter energy density, i.e., we will
treat the dark matter and baryons as a single matter fluid.  One
should also recall that baryons total about one-sixth of the present
total matter distribution, thus, we expected that by not making a
distinction  between dark matter and baryons, we should not be
incurring in any significant error in the  numerical evaluations in
the next sections. This should, in particular, be true at least at the
background level, which is the case we focus on this work.

As far as dark matter production is concerned, WI itself can have the
mechanisms needed for generating it. This can happen either through
the intrinsic dissipative dynamics inherent of the WI picture, or also
as a consequence of it, like, e.g., the inflaton itself being a stable
remnant behaving at later times as  dark
matter~\cite{Bastero-Gil:2015lga,Rosa:2018iff}. 
{}For some decaying vacuum energy models that can possibly behave like in WI, see,
e.g.,~Ref.~\cite{Dymnikova:1998ps}.
The possibility of
matter generation (through a baryogenesis scenario) in WI was explored
before in Refs.~\cite{Brandenberger:2003kc,BasteroGil:2011cx}. In
particular, iRef.~\cite{Bastero-Gil:2014oga} studied in
detail the observational implications of this scenario, with the
prediction that the matter production in WI would lead to fully
anticorrelated isocurvature perturbations. These previous studies
motivate us to treat the matter evolution dynamics as concomitant with
those from the scalar field and the radiation fluid.  Indeed, the
dissipation model studied in
Refs.~\cite{Brandenberger:2003kc,Bastero-Gil:2014oga} was actually the
same that leads to the dissipation coefficient of the form of
Eq.~(\ref{cubic}).  The recent models of dark matter and dark energy
interactions~\cite{Bolotin:2013jpa,Wang:2016lxa} also motivate us to
consider similar energy exchange terms involving the dark
sector. Here, however, we motivate these terms from the WI
picture. Indeed, we can show that by simply replacing the radiation
energy density by the matter density, $T \propto \rho_R^{1/4} \to
\rho_m^{1/4}$, we can construct dissipation ratios $Q_{\rho_m}$ with
dynamical behavior similar to the ones seen in Eq.~(\ref{dQdNexpn}).
We will use this generic property in our construction below.
Motivated by the above remarks, here we propose the complete
set of background equations involving the quintessential scalar field,
the radiation fluid and the matter energy densities as given,
respectively, by
\begin{eqnarray}
&& \ddot \phi + 3 \left(1+Q \right) H \dot \phi  + \Upsilon_{\rho_m}
  \dot \phi + V_{,\phi}=0,
\label{eqphinew}
\\ && \dot \rho_R + 4 H \rho_R = 3H Q\dot \phi^2 ,
\label{eqrhoRnew}
\\ && \dot \rho_m + 3 H \rho_m = \Upsilon_{\rho_m} \dot \phi^2,
\label{eqrhom}
\end{eqnarray}
where $\Upsilon_{\rho_m}$ describes the energy exchange term between
the quintessential scalar field and matter energy density and we
parametrize it in the form
\begin{eqnarray}
\Upsilon_{\rho_m} = c_m \frac{\rho_m^{3/4}}{\phi^2} + 
\frac{M^2}{\rho_m^{1/4}},
\label{Upsilonm}
\end{eqnarray}
where $c_m$ is a dimensionless constant and $M$ is another constant with mass
dimension. The form of the first term in Eq.~(\ref{Upsilonm}) is closely
motivated by the Eq.~(\ref{cubic}). Similarly to the behavior found
for the dissipation coefficient of the form Eq.~(\ref{cubic}), the
first term in Eq.~(\ref{Upsilonm}) displays a growing dynamics during
inflation, setting an abundance for the matter density until around the
beginning of the radiation-domination regime, after which it decays
with time. On the other hand,  the second term in Eq.~(\ref{Upsilonm})
displays opposite behavior, decreasing throughout inflation but
later on growing with time. Given appropriate parameters $c_m$ and
$M$, we can arrange a similar behavior found, e.g., in the case of
nonminimal couplings of the scalar field  to
matter~\cite{Amendola:1999er}.  With the second term in
Eq.~(\ref{Upsilonm}), as it grows at later times, it will eventually
provide an extra friction force on the quintessence scalar field and
help making $\phi$ acquire a negative equation of state,
signaling the beginning of the dark energy (quintessence) domination
epoch. 
As far the dependences on $\rho_m$ in Eq.~(\ref{Upsilonm}) are concerned,
we could also think in a term similar to a dissipation coefficient that
is proportional to the temperature in the WI scenario~\cite{Bastero-Gil:2016qru}.
Then, by the same analogy as assumed in the construction of the terms in Eq.~(\ref{Upsilonm}),
by taking $T \to \rho_m^{1/4}$, such a term would have a behavior similar to
the second term on the right-hand side of Eq.~(\ref{Upsilonm}), growing at the end of
the inflationary regime and leading the energy density of the scalar field
to dominate at present. However, the growth is much slower than provided by the
second term in Eq.~(\ref{Upsilonm}), making the transition from matter domination to
the acceleration regime happen at a much higher redshift than is observationally
acceptable (likewise, the matter domination regime is too short).
Such a term in Eq.~(\ref{Upsilonm}) is then rendered unacceptable. Likewise,
we could look at the first term in Eq.~(\ref{Upsilonm}) with a larger power in the
matter energy density, but then we face the opposite behavior, a potential larger
matter domination regime and much later transition to the acceleration regime
(and also possibly instabilities issues as far as a dynamical system analysis
is concerned, similar to what is seen in WI with dissipation coefficients
with a large power in the temperature~\cite{Moss:2008yb,delCampo:2010by,BasteroGil:2012zr}). 
In this sense, the dependences in Eq.~(\ref{Upsilonm}) seem a middle term between these
extreme behaviors.

In the next sections, we will show explicitly how each of the dissipative
terms appearing in Eqs.~(\ref{eqphinew}), (\ref{eqrhoRnew}) and 
(\ref{eqrhom}), and given by Eqs.~(\ref{cubic}) and (\ref{Upsilonm}), help 
produce a consistent picture for the evolution of the
scalar (quintessential inflaton) field and the radiation and matter
constituents, from the primordial inflation time up to the recent
epoch. We will start by first studying the inflationary early Universe
epoch.

%%%%%%%%%%%%%%%%%%%%%%%%%%%%%%%%%%%%%%%%%%%%%%%%%%%%%%%%%%%%%%
\section{Early Universe inflationary cosmological dynamics}
\label{sec3}

The potential (\ref{pot}) can support an inflationary regime in its
flat region, $\alpha (\phi/M_{\rm Pl})^n \ll 1$. The predictions for
this model can be contrasted  with those from the recent CMB
observations, e.g., from the Planck data~\cite{Akrami:2018odb}.  We
are, in particular, interested in the tensor-to-scalar ratio $r$ and the
spectral tilt $n_s$, defined, respectively, by
\begin{equation}
r= \frac{\Delta_{T}}{\Delta_{\cal R}},
\label{eq:r}
\end{equation}
and
\begin{equation}
n_s -1 = \lim_{k\to k_*}   \frac{d \ln \Delta_{{\cal R}}(k/k_*) }{d
  \ln(k/k_*) },
\label{eq:n}
\end{equation}
where $\Delta_{\cal R}$ is the primordial scalar curvature power
spectrum and $\Delta_{T}$ is the tensor power spectrum.  Quantities with
a subindex $*$ mean that they are evaluated at the Hubble radius
crossing,  $k_*=a_* H_*$.  The Planck
Collaboration~\cite{Akrami:2018odb} gives for $r$ the upper bound,
$r<0.056$ (95$\%$ CL, Planck TT,TE,EE+lowE+lensing+BK15, at the pivot
scale $k_p = 0.002/{\rm Mpc}$), while for the spectral tilt the result
is $n_s=0.9658\pm 0.0040$  (95$\%$ CL, Planck
TT,TE,EE+lowE+lensing+BK15+BAO+running).  {}Furthermore, the
normalization of the primordial scalar curvature power spectrum, at
the pivot scale $k_*$, is given by $\ln\left(10^{10} \Delta_{{\cal R}}
\right) \simeq 3.047$ (TT,TE,EE-lowE+lensing+BAO 68$\%$ limits),
according to the Planck Collaboration~\cite{Aghanim:2018eyx} and this
is the value we will assume in all our numerical simulations, in
particular for finding the normalization $V_0$ of the potential
Eq.~(\ref{pot}).

In the cold inflation scenario, i.e., in the absence of dissipative
effects and no radiation bath during inflation, $\Delta_{{\cal R}}$
and $\Delta_{T}$ are given, respectively, by the standard
expressions~\cite{lyth2009primordial}
\begin{eqnarray} \label{PkCI}
\Delta_{{\cal R}} &=&  \left(\frac{ H^2}{2 \pi\dot{\phi}}\right)^2 ,
\\ \Delta_{T} &=& \frac{2 H^2}{\pi^2 M_{\rm Pl}^2}.
\label{tensor}
\end{eqnarray}

%%%%%%%%%%%%%%FIGURE01%%%%%%%%%%%%%%%%%%%
\begin{center}
\begin{figure}[!htb]
\includegraphics[width=8.6cm]{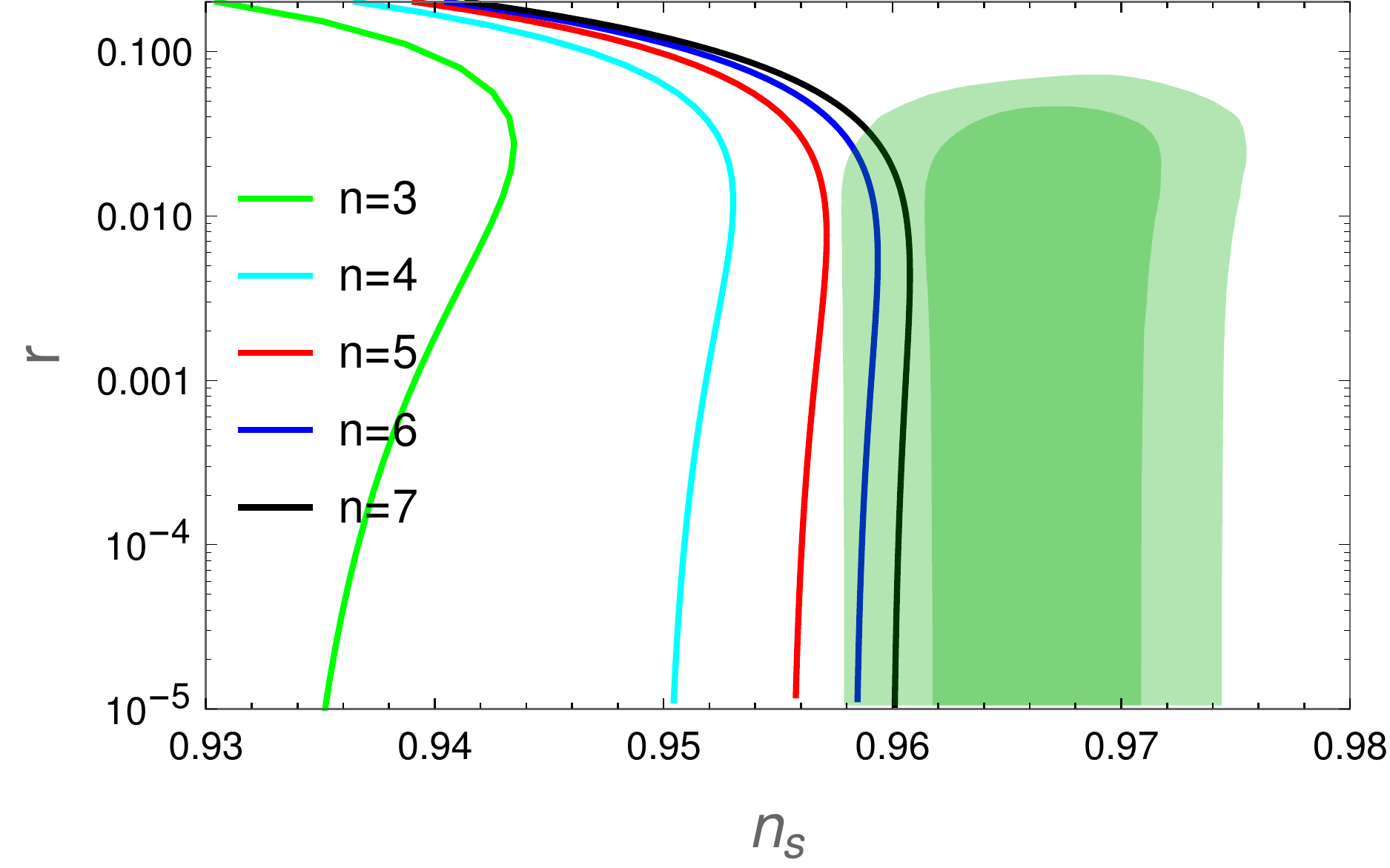}
\caption{The spectral index $n_s$ and the tensor-to-scalar ratio $r$
  in the plane $(n_s,r)$, in the cold inflation scenario (i.e., in the
  absence of dissipative effects)  for different values of the
  exponent $n$ and constant  $\alpha$, such that $10^{-12} \leq \alpha
  \leq 10^{-1}$ ($\alpha$ increases when going from small to larger
  values of $r$).  The shaded areas are for the $68\%$ and $95\%$
  C.L. results from Planck 2018 (TT+TE+EE+lowE+lensing+BK15+BAO
  data). }
\label{fignsXrCI}
\end{figure}
\end{center}
%%%%%%%%%%%%%%%%%%%%%%%%%%%%%%%%%%%%%%%%

In the Ref.~\cite{Geng:2017mic}, the authors have studied in detail
the predictions for the model with the potential given by
Eq.~(\ref{pot}), including approximate analytical expressions for $r$
and $n_s$ in the cold inflation case. {}For completeness, in
{}Fig.~\ref{fignsXrCI} we show these results for different values of
the power $n$ in Eq.~(\ref{pot}).  The dimensionless constant $\alpha$
in the potential varies, $10^{-12} \leq \alpha \leq
10^{-1}$.  The values of $\alpha$ increase when going from small to
larger values of $r$, which then determines the curves in
{}Fig.~\ref{fignsXrCI}. {}For this figure, we have considered $N_*=60$
for the number of e-folds before the end of inflation.

One notices from {}Fig.~\ref{fignsXrCI} that only for $n>5$ can one
have results for the spectral tilt compatible with the recent
observational data at the 2$\sigma$ level. This situation can change
considerably in the WI case.  Because of dissipation and the presence
of a radiation bath, the primordial scalar power spectrum  given by
Eq.~(\ref{PkCI}) changes, while the tensor spectrum Eq.~(\ref{tensor})
remains unchanged.  The primordial power  spectrum for WI at horizon
crossing can be expressed in the form (see, e.g.,
Refs.~\cite{Ramos:2013nsa,Bartrum:2013fia,Benetti:2016jhf})
\begin{equation} \label{Pk}
\Delta_{{\cal R}}(k/k_*) =  \left(\frac{ H_{*}^2}{2
  \pi\dot{\phi}_*}\right)^2  {\cal F} (k/k_*),
\end{equation}
where the function ${\cal F} (k/k_*)$ in Eq.~(\ref{Pk})  is given by
\begin{equation}
{\cal F} (k/k_*) \equiv  \left(1+2n_* + \frac{2\sqrt{3}\pi
  Q_*}{\sqrt{3+4\pi Q_*}}{T_*\over H_*}\right) G(Q_*),
\label{calF}
\end{equation}
where $n_*$ denotes the inflaton statistical distribution due to the
presence of the radiation bath and $G(Q_*)$ accounts for the effect of
the coupling of the inflaton and radiation
fluctuations~\cite{Graham:2009bf,BasteroGil:2011xd,Bastero-Gil:2014jsa}.
$G(Q_*)$, in general, can only be determined by numerically solving the
set of perturbation equations in WI. However, in the weak dissipative
regime of WI, where $Q_* \ll 1$, these coupling effects on the scalar
curvature power spectrum are negligible and we can safely consider
$G(Q_*) \simeq 1$ in Eq.~(\ref{calF}). In the present work we will
restrict to this regime, which will suffice for us to obtain our main
results. {}For a note on this point of considering the weak dissipative
regime, one should, however, note that evading the recent so-called
Swampland conjectures, one  typically requires going in the strong
dissipative regime of WI, $Q \gg 1$, as shown in
Refs.~\cite{Motaharfar:2018zyb,Bastero-Gil:2019gao}. With the
generalized exponential potential Eq.~(\ref{pot}) we can find
parameter regimes supporting  $Q \gg 1$, but in the present work we
will not explore this region of parameters but leave this open for
future studies. In fact, even the simple exponential potential (with
$n=1$) is also shown to support the strong dissipative regime and also
lead to consistent inflationary observables, provided that we extend
the system beyond the general relativity case, e.g., in the context of
extra dimensions and the braneworld extension, as shown recently in
Ref.~\cite{Kamali:2019xnt}, in which all the swampland conjectures can be
overcome. 

Before presenting our results for the WI case, let us comment
on the choice of the initial conditions in our problem. Deep in the inflationary 
regime, we can use the slow-roll equations derived from Eqs.~(\ref{eqphinew})
and (\ref{eqrhoRnew}). The matter energy density is negligible during that time and can
be neglected (including $\Upsilon_{\rho_m}$). With these equations and also using 
the expression for the scalar curvature power spectrum Eq.~(\ref{Pk}) and the corresponding 
CMB normalization, we can fix the normalization $V_0$ in the inflaton potential and find 
the appropriate initial conditions leading to the required $N_*$ e-folds of inflation
before its end, which is the point in the inflationary evolution relevant for obtaining
$r$ and $n_s$.
In particular, the value for $\phi_*$ obtained
is always a bit farther away to the right from the inflection point of the potential 
(when $n$ is odd), or away (also to the right) from the top of the potential hill 
(when $n$ is even), since otherwise we could be starting 
with initial conditions leading to a much larger value of e-folds of inflation and having to wait some
long time until the relevant instant $N_*$ when the scales of interest leave the Hubble radius. 
Here, as also in many other works on inflation, including Ref.~\cite{Geng:2017mic}, which
made explicit studies of the present generalized exponential inflaton potential, we do not address
how exactly the system achieves the conditions necessary for inflation.
Physically viable suggestions for mechanisms setting the appropriate
conditions for inflation are, for example, through some earlier phase transition, 
or through thermal and dissipative effects acting in the preinflationary phase, 
as suggested and developed in Ref.~\cite{Bastero-Gil:2016mrl}.

%%%%%%%%%%%%%%FIGURE02%%%%%%%%%%%%%%%%%%%
\begin{center}
\begin{figure}[!htb]
\includegraphics[width=8.6cm]{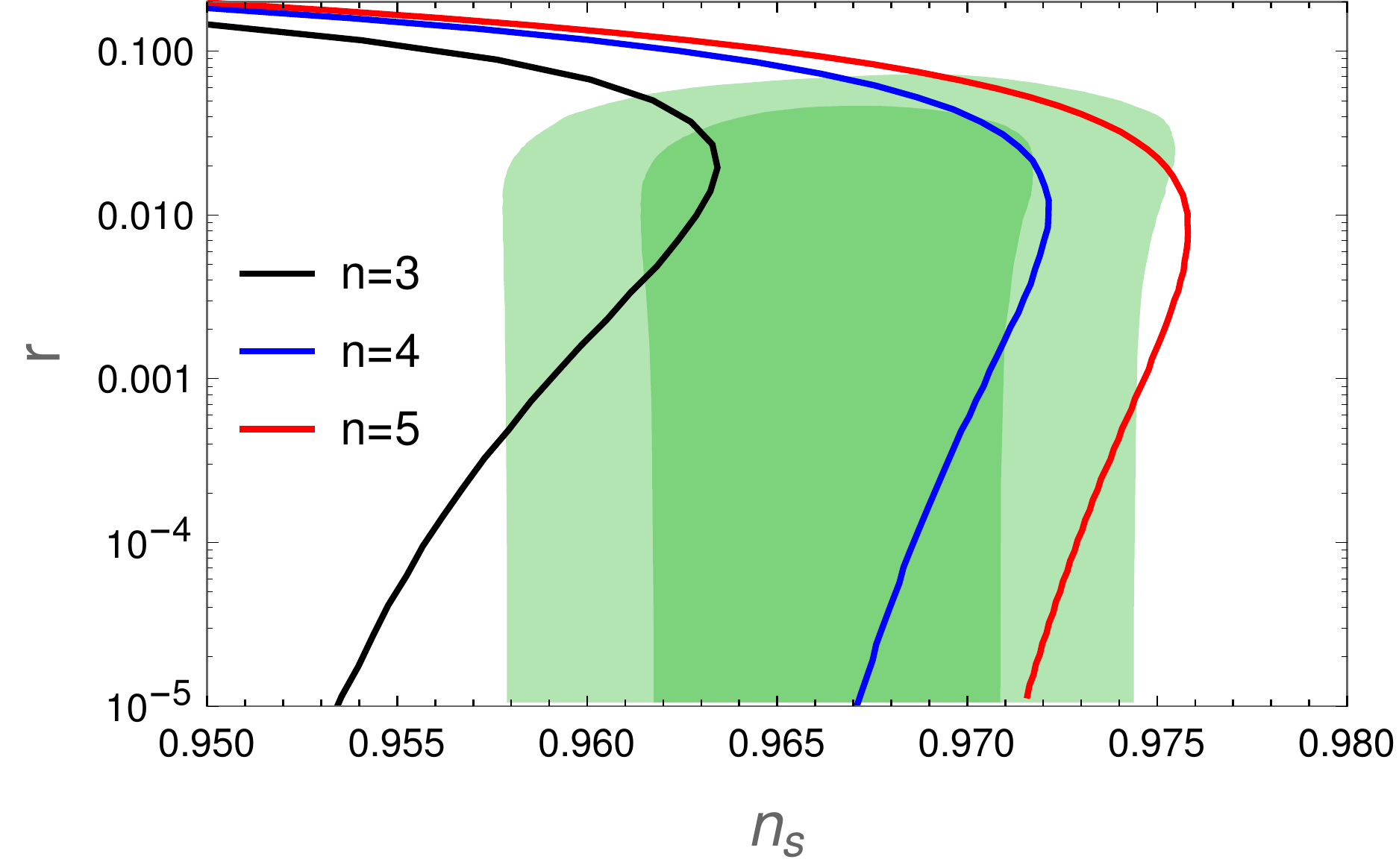}
\caption{Similar to {}Fig.~\ref{fignsXrCI}, but in the WI scenario,
  for a  dissipation ratio $Q_*=5 \times 10^{-7}$. Only the
  $n=3,\,4,\,5$ cases are shown. }
\label{fignsXrWI}
\end{figure}
\end{center}
%%%%%%%%%%%%%%%%%%%%%%%%%%%%%%%%%%%%%%%%

By performing the analysis in the WI case, with scalar power spectrum
given by Eq.~(\ref{Pk}) and setting $G(Q_*)=1$ in Eq.~(\ref{calF}),
which is justified for $Q_* \ll 1$, we show in {}Fig.~\ref{fignsXrWI}
the results equivalent to  {}Fig.~\ref{fignsXrCI}.  In
{}Fig.~\ref{fignsXrWI} we have fixed the dissipation ratio at the
value $Q_*=5\times 10^{-7}$. Though very small, it is  already large
enough to deform the curves (with varying $\alpha$) shown in the
previous {}Fig.~\ref{fignsXrCI}. In particular, the models with $n=3$
and $n=4$, which were excluded beforein the cold inflation scenario,
are now found to be consistent with the Planck data for a range of
$\alpha$ values\footnote{However, we find no values for the
  dissipation  and coefficient $\alpha$ such to get the case with
  $n=2$ inside the Planck confidence regions, which has a $n_s$ that
  is always too red tilted.}.

\begin{center}
\begin{figure}[!htb]
\includegraphics[width=8.6cm]{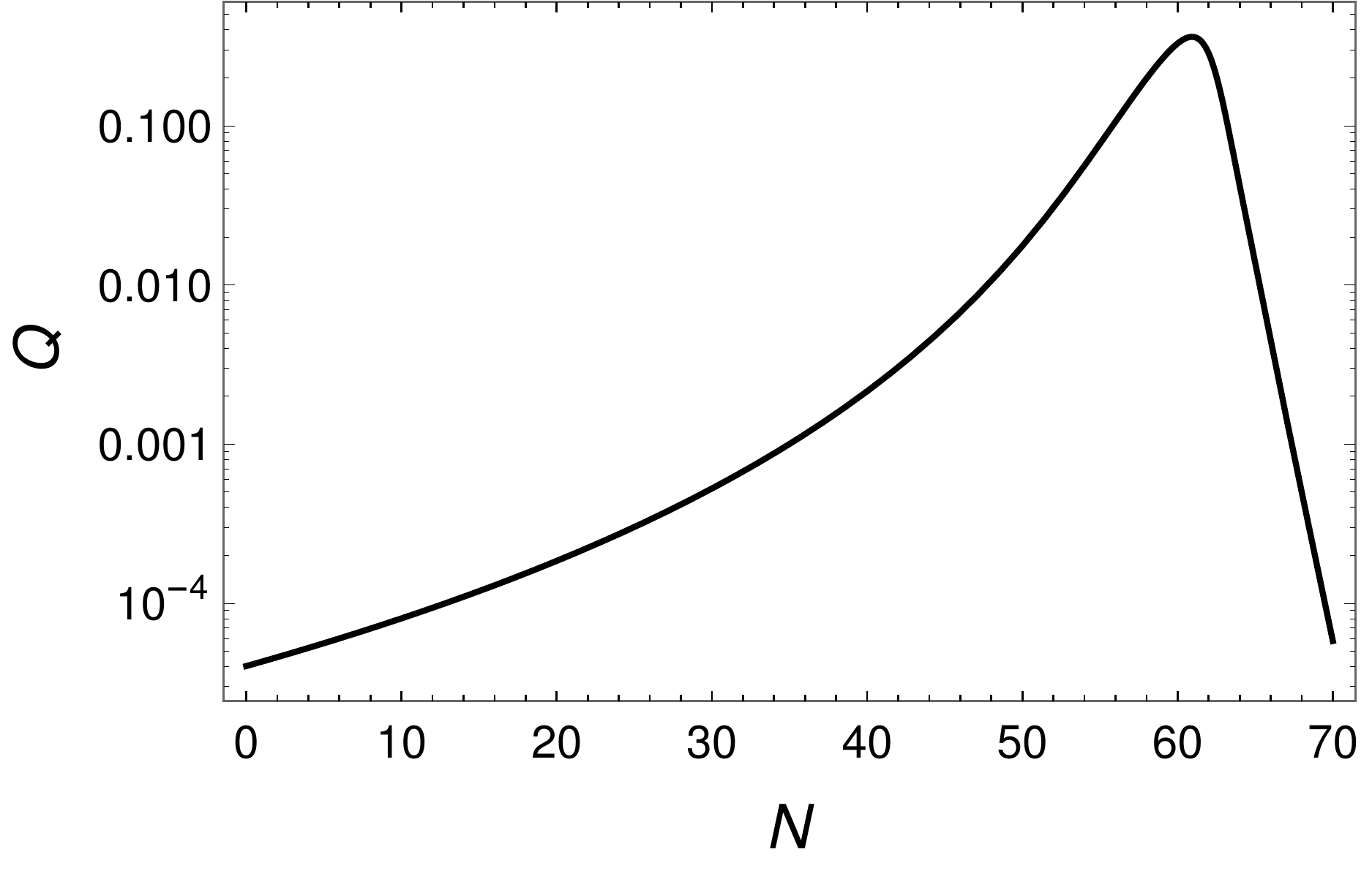}
\caption{The evolution of the dissipation ratio $Q$ with the number of
e-folds for the model $n=3$ and parameters given in Table~\ref{tab1}. }
\label{figQ}
\end{figure}
\end{center}

We also recall from Eq.~(\ref{dQdNexpn}) that $Q$ is evolving with the
number of e-folds. Thus, even though we might have a very small $Q_*$ at the
instant $N_*$ e-folds before the end of inflation (and still also
be inside the WI regime, where $T> H$), by the end of inflation,
and before the scaling regime for which $\alpha (\phi/M_{\rm Pl})^n \gtrsim 1$,
$Q_{\rm end}$ can be much larger than $Q_*$. This is in fact what facilitates 
the transition from the inflationary regime to the radiation-dominated one.
We illustrate this behavior in {}Fig.~\ref{figQ}, taking as an illustration, the
case with $n=3$ in Eq.~(\ref{pot}) and for a dissipation ratio $Q_*=6.24\times 10^{-6}$,
which is one of the cases explicitly studied in the next section (see
Table~\ref{tab1}).

%%%%%%%%%%%%%%FIGURE03%%%%%%%%%%%%%%%%%%%
%%%%%%%%%%%%%%%%%%%%%%%%%%%%%%%%%%%%%%%%

\begin{center}
\begin{figure}[!htb]
\includegraphics[width=8.6cm]{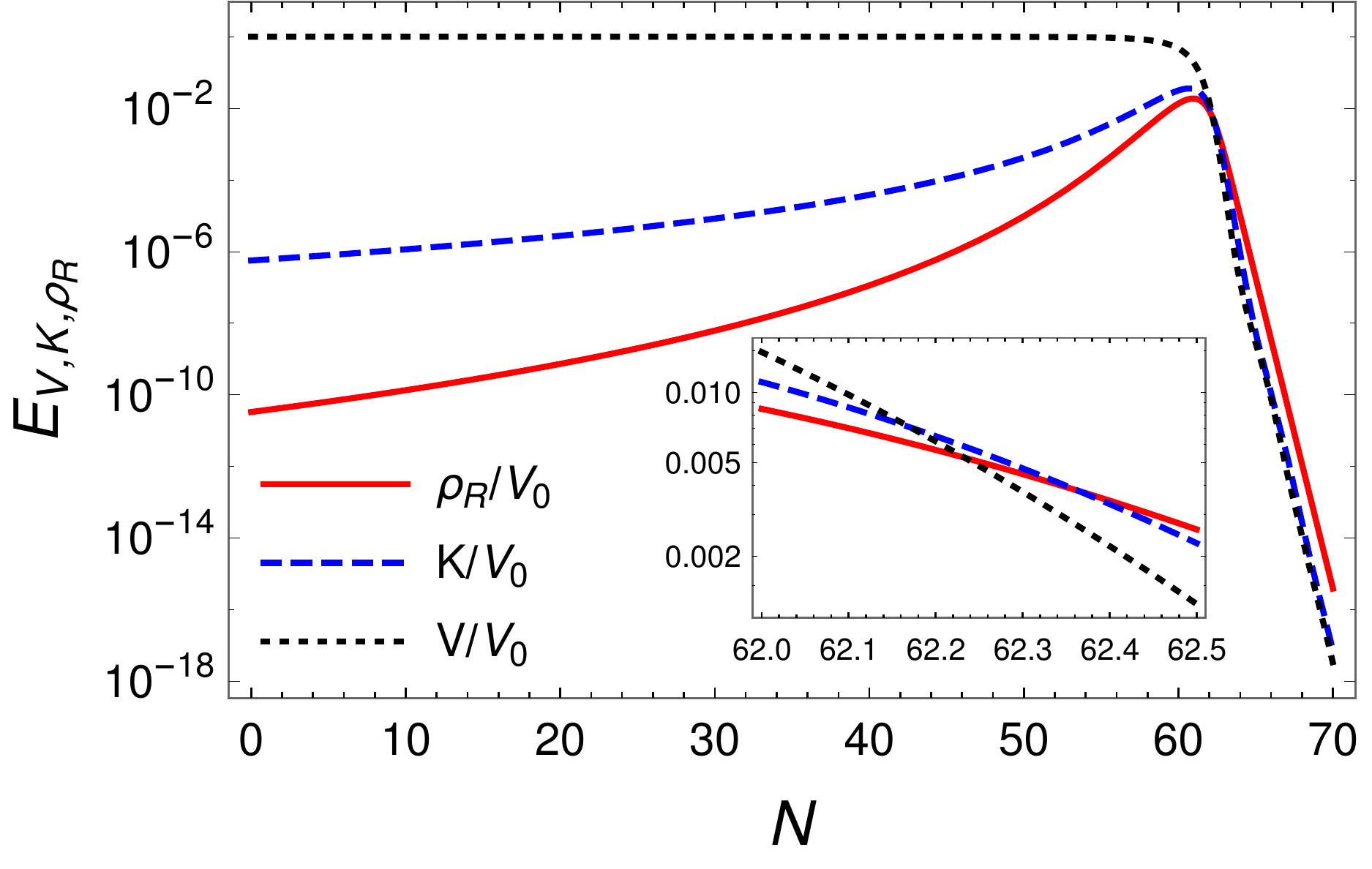}
\caption{The evolution of the kinetic 
$K={\dot \phi}^2/2$,  of the potential $V(\phi)$, and for the
radiation $\rho_R$ energy densities, for the same model and parameters used in {}Fig.~\ref{figQ}.
All energy densities are normalized by $V_0$, where $V_0\simeq 1.62 \times 10^{15}{\rm GeV}$
for the present model parameters.
The inset zooms in a region right after inflation.}
\label{figenergies}
\end{figure}
\end{center}

In the model and parameters considered in {}Fig.~\ref{figQ}, we have
$N_* \simeq 61.6$. We see that during inflation $Q$ grows by more than 4
orders of magnitude and at the end of inflation, $Q_{\rm end} \simeq 0.33$.
Soon after that and as anticipated in the discussion following Eq.~(\ref{dQdNexpn}),
with the scalar field satisfying $\alpha (\phi/M_{\rm Pl})^n \gtrsim 1$,
the value of $Q$ quickly drops down.
It is also useful to show the dynamics for the inflaton kinetic 
$K={\dot \phi}^2/2$,  the potential  $V(\phi)$, and the
radiation $\rho_R$ energy densities. This is shown in {}Fig.~\ref{figenergies}.

%%%%%%%%%%%%%%FIGURE04%%%%%%%%%%%%%%%%%%%
%%%%%%%%%%%%%%%%%%%%%%%%%%%%%%%%%%%%%%%%

We notice from the results shown in {}Fig.~\ref{figenergies} that before the
system transits to the radiation-dominated phase, there is a small regime
in which the kinetic energy of the scalar field is the largest energy component.
This kinetic regime lasts however a very short time, however, about $\sim 0.2$ e-folds
in the present example. This is a characteristics of the WI dynamics
and also seen in the warm quintessential inflation studied e.g. in Ref.~\cite{Dimopoulos:2019gpz}.

Having established that one can have a consistent inflationary regime
in the WI dynamics with the primordial steep potentials of the form of
Eq.~(\ref{pot}), let us now analyze the late-time dynamics for the
model, with background equations given by Eqs.~(\ref{eqphinew}),
(\ref{eqrhoRnew}) and (\ref{eqrhom}), with dissipation coefficients
given by Eqs.~(\ref{cubic}) and (\ref{Upsilonm}). We will focus on
three models in particular, in which $n=3,\,4,\,5$.

%%%%%%%%%%%%%%%%%%%%%%%%%%%%%%%%%%%%%%%%%%%%%%%%%%%%%%%%%%%%%%
\section{Late-time Universe dynamics}
\label{sec4}

{}For the late-time dynamics, we will work with three explicit
examples,  considering the cases of $n=3,\,4,\,5$ in the potential
Eq.~(\ref{pot}).  The constant $\alpha$ in the potential and the
dissipation ratio $Q_*$ determining the inflationary dynamics,  
such as the value for spectral tilt $n_s$ that we find is
always close to the central value from the Planck data, i.e., $n_s
\simeq 0.965$, are chosen. 
The number of e-folds of inflation $N_*$ between the moment the
relevant  scales with wave number $k_*$ leave the Hubble radius and
reenter around today, is defined by the relation~\cite{Liddle:2003as}
\begin{equation}
\frac{k_*}{a_0 H_0} = e^{-N_*} \frac{T_0}{T_{\rm end}}
\frac{H_*}{H_0},
\label{N*}
\end{equation}
where $a_0$ and $H_0$ are the today values for the scale factor and
Hubble parameter, respectively.  $H_*$ is the Hubble parameter during
inflation and $T_{\rm end}$ and $T_0$ are the temperature at the
beginning of the radiation-dominated regime and the CMB temperature
today, respectively. Typically, in the cold inflation scenario $T_{\rm
  end}$ is difficult in  general to obtain given that it depends of
the details of the reheating process after inflation.  In WI, however,
this is much simplified, since the transition from the end of
inflation to the radiation-dominated regime is smooth and $T_{\rm
  end}$ is simply the temperature at the end of WI.  Therefore, we can
self-consistently obtain $N_*$ from Eq.~(\ref{N*}).

In this work, we take the convention that $a_0=1$ (likewise, the
redshift parameter today is $z_0=0$). {}For the Hubble parameter
today,  we will assume the Planck result, $H_0=67.66\, {\rm km}\,
s^{-1} {\rm Mpc}^{-1}$ [from the Planck
Collaboration~\cite{Aghanim:2018eyx}, TT,TE,EE-lowE+lensing+BAO 68$\%$
limits,  $H_0 = (67.66 \pm 0.42)\, {\rm km}\, s^{-1} {\rm Mpc}^{-1}$].
Likewise, for the CMB temperature today we assume the value $T_0 =
2.725\, {\rm K}= 2.349 \times 10^{-13}\, {\rm GeV}$. The value of $T_0$
will fix the total number of e-folds of expansion, $N_{\rm total}$
from the instant $N_*$ deep in the inflationary regime until today.
We still need to find the constants $c_m$ and
$M$ in Eq.~(\ref{Upsilonm}).  These can be unambiguously found by
requiring that we obtain the appropriate values for the dark matter and
dark energy density ratios today, $\Omega_{m,0}$ and $\Omega_{{\rm
    DE},0}$, respectively.  {}For these values, we use again the 
    dataset TT,TE,EE-lowE+lensing+BAO 68$\%$ limits from the Planck
Collaboration, which give the values  $\Omega_{m,0}=0.3111\pm 0.0056$
and  $\Omega_{{\rm DE},0}=0.6889\pm 0.0056$ (see, e.g., Table 2 of
Ref.~\cite{Aghanim:2018eyx}).  Our model parameters are determined
such as to lead to values for $\Omega_{m,0}$ and $\Omega_{\phi,0}$
that are between the range found in these observational results, which
are here taken to be our fiducial values. Likewise, the fractional density 
in radiation, $\Omega_r$, is always at a level consistent with its value
today, $\Omega_{r,0} \sim 8 \times 10^{-5}$ (including neutrinos).

\begin{widetext}
%%%%%%%%%%%%%%%%%%%%%%%%%%%%%%%%%%%%%%%%%%%%%%%%%%%%%%%%%%%%%
\begin{center}
\begin{table}[!htb]
\caption{The three model examples considered, along with the respective parameters
and the relevant cosmological quantities (see text) obtained from them.}
\label{tab1}
\begin{tabular}{c|c|c|c|c|c|c|c|c}
\hline
\hline
Model & $n_s$ & $r$ & $N_*$ & $N_{\rm total}$ & $c_m$ & $M$ (eV) & $z_{\rm DM-DE}$ & $\Omega_\phi(z_{\rm rec} \approx 1100)$\\
\hline
$n=3$ &  &  &  &  &  &  &  & \\
$\alpha= 0.05$& 0.9652 & $6.24\times 10^{-6}$ & 61.6 & 125.2 & 0.0668 & $1.48\times 10^{-16}$ & 0.6135 & 0.00186\\
$Q_*=4.03\times 10^{-5}$ &  &  &  &  &  &  &  & \\
\hline
$n=4$ &  &  &  &  &  &  &  & \\
$\alpha=0.01$ & 0.9659 & $9.24\times 10^{-5}$ & 62.1 & 126.6 & 0.0455 & $2.72\times 10^{-16}$ & 0.6051 & 0.00056\\
$Q_*= 2.55\times 10^{-7}$ &  &  &  &  &  &  &  & \\
\hline
$n=5$ &  &  &  &  &  &  &  & \\
$\alpha=4\times 10^{-4}$ & 0.9657 & $4.24 \times 10^{-4}$ & 62.6 & 127.3 & 0.0723 & $2.99\times 10^{-16}$ & 0.6050 & 0.00047\\
$Q_*= 8.73\times 10^{-8}$ &  &  &  &  &  &  &  & \\
\hline
\hline
\end{tabular}
\end{table}
\end{center}
%%%%%%%%%%%%%%%%%%%%%%%%%%%%%%%%%%%%%%%%%%%%%%%%%%%%%%%%%%%%%%%%%%%%%%%%%%
\end{widetext}

Our results for the three examples using  $n=3,\,4,\,5$ are summarized
in Table~\ref{tab1}.  
We also quote the results obtained for the redshift at the transition
from dark matter to dark energy domination, $z_{\rm DM-DE}$. In
Table~\ref{tab1} we also give the value for the density parameter for
the quintessential scalar field close to recombination,
$\Omega_\phi(z_{\rm rec} \approx 1100)$, which is a useful result to
be compared with early dark energy constraints.

The values obtained for $M$ and given in  Table~\ref{tab1}, interestingly,
they fall in the ballpark for the masses of light axionlike
pressureless cold dark matter particles~\cite{Amendola:2016saw}. 
In principle we could also directly write $M$ in terms of some typical 
mass scale for dark matter, up to some constant $c_M$, e.g.,  $m_{\rm DM} = c_M M$.  
The two dissipation terms in Eq.~(\ref{Upsilonm}) in principle should both come
from similar physical processes, thus, the constants
$c_m$ and $c_M$ are not expected to differ in value by too disparate numbers.
Even when setting, for example, $c_M \ll c_m$, $m_{\rm DM}$ can reach the mass 
range of the so-called fuzzy cold dark matter~\cite{Hu:2000ke}, 
$10^{-33} {\rm eV}
\lesssim m_{\rm DM} \lesssim 10^{-18}{\rm eV}$. Even though we did not
make any assumptions here about the underline dark matter being
produced (except that which is cold), the values of $M$ obtained and
shown  in  Table~\ref{tab1} are quite suggestive.
Despite that the mass scale $M$ might have some physically well-motivated
value, its value is, in principle, not related to any other natural energy
scale in our model. Thus, its smallness can still be associated with some level of fine-tuning
due to $M$ being much smaller than the vacuum energy density in the 
$\Lambda$CDM model, $\rho_\Lambda^{1/4} \sim 10^{-3} {\rm eV}$, but 
still much larger than $H_0\sim 10^{-33}{\rm eV}$.

Early dark energy can potentially change the expansion history in the
early and also in the late Universe.  It can affect galaxy formation,
big bang nucleosynthesis and the CMB anisotropies.  The strongest
bound to date come from the CMB data~\cite{Ade:2015rim}, which sets
the upper bound  $\Omega_\phi(z_{\rm rec}) \lesssim 0.0036$ (at 95$\%$
confidence level, for Planck TT,TE,EE-lowP+BSH) on the amount of dark
energy (here described by quintessence) at the time of recombination,
$z_{\rm rec} \approx 1100$.  Our results for $\Omega_\phi(z_{\rm
  rec})$ for the three models analyzed and given in Table~\ref{tab1}
are shown to satisfy this upper bound.

The redshift at the transition from dark matter to dark energy
domination, $z_{\rm DM-DE}$, has been considered as a possible
additional cosmological parameter. Though it is very difficult to precisely
constrain $z_{\rm DM-DE}$ (see, e.g., Ref.~\cite{Jesus:2019nnk} for a recent
analysis), it is generically assumed that any model describing dark
energy and the current cosmological data, should have a value such
that $0.5 < z_{\rm DM-DE} < 1$. Again, our results for the three
models analyzed produce acceptable values for  $z_{\rm DM-DE}$, as
shown  in Table~\ref{tab1}.

%%%%%%%%%%%%%%FIGURE05%%%%%%%%%%%%%%%%%%%
\begin{center}
\begin{figure}[!htb]
\includegraphics[width=8.6cm]{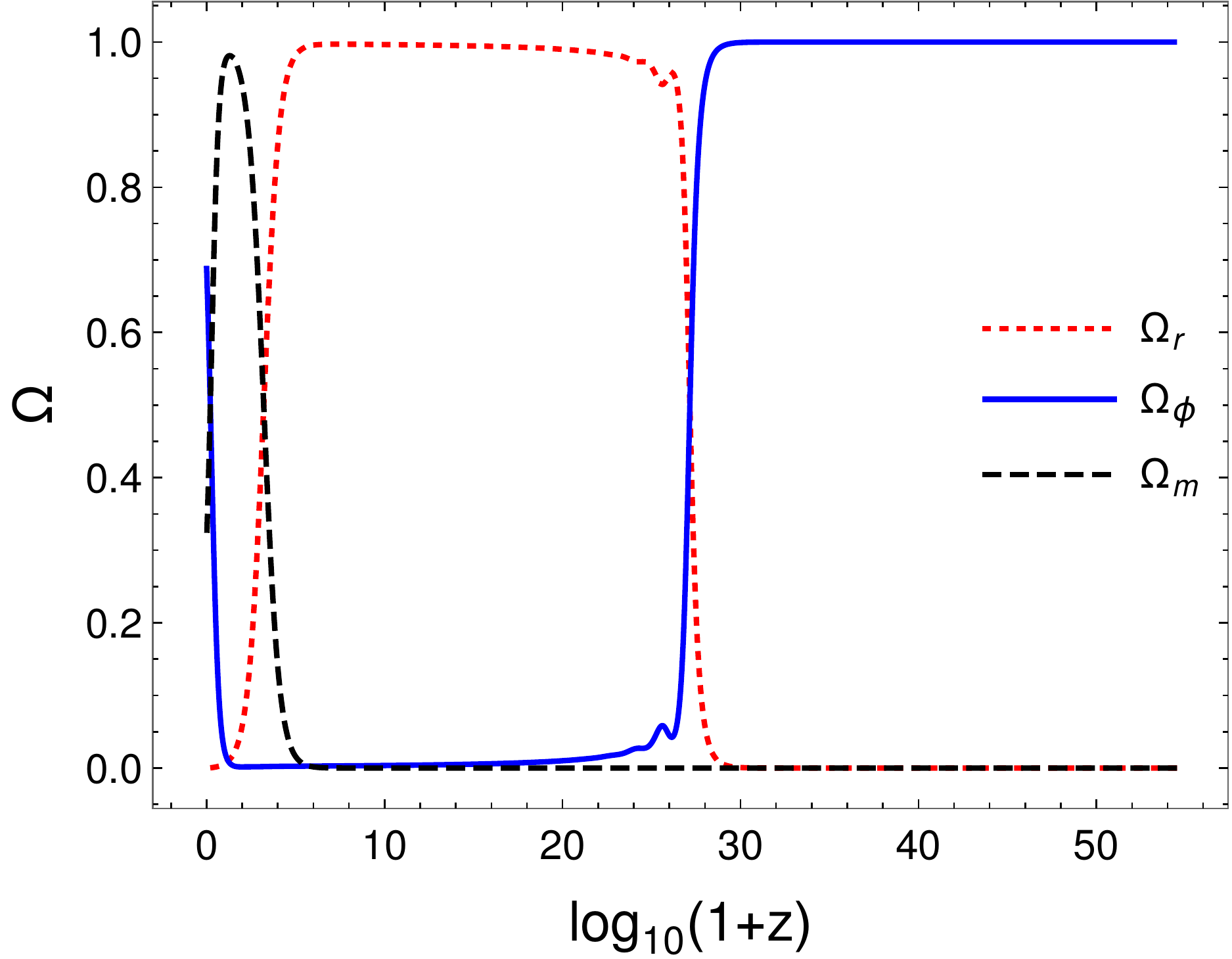}
\caption{The energy density fractions,  $\Omega_\phi,\, \Omega_m$ and
  $\Omega_{r}$, as a function of the redshift, for the model $n=3$ and
  parameters given in Table~\ref{tab1}. }
\label{figOmegas}
\end{figure}
\end{center}
%%%%%%%%%%%%%%%%%%%%%%%%%%%%%%%%%%%%%%%%

In {}Fig.~\ref{figOmegas}, we show the complete evolution for each of
the energy density fractions,  $\Omega_\phi,\, \Omega_m$ and
$\Omega_{r}$, for the quintessence scalar field, the matter and
radiation fluid components.  This is shown for the model with $n=3$
with the parameters given in Table~\ref{tab1}. The other two models
have very similar results and, hence, we do not show them here.

%%%%%%%%%%%%%%FIGURE06%%%%%%%%%%%%%%%%%%%
\begin{center}
\begin{figure}[!htb]
\includegraphics[width=8.6cm]{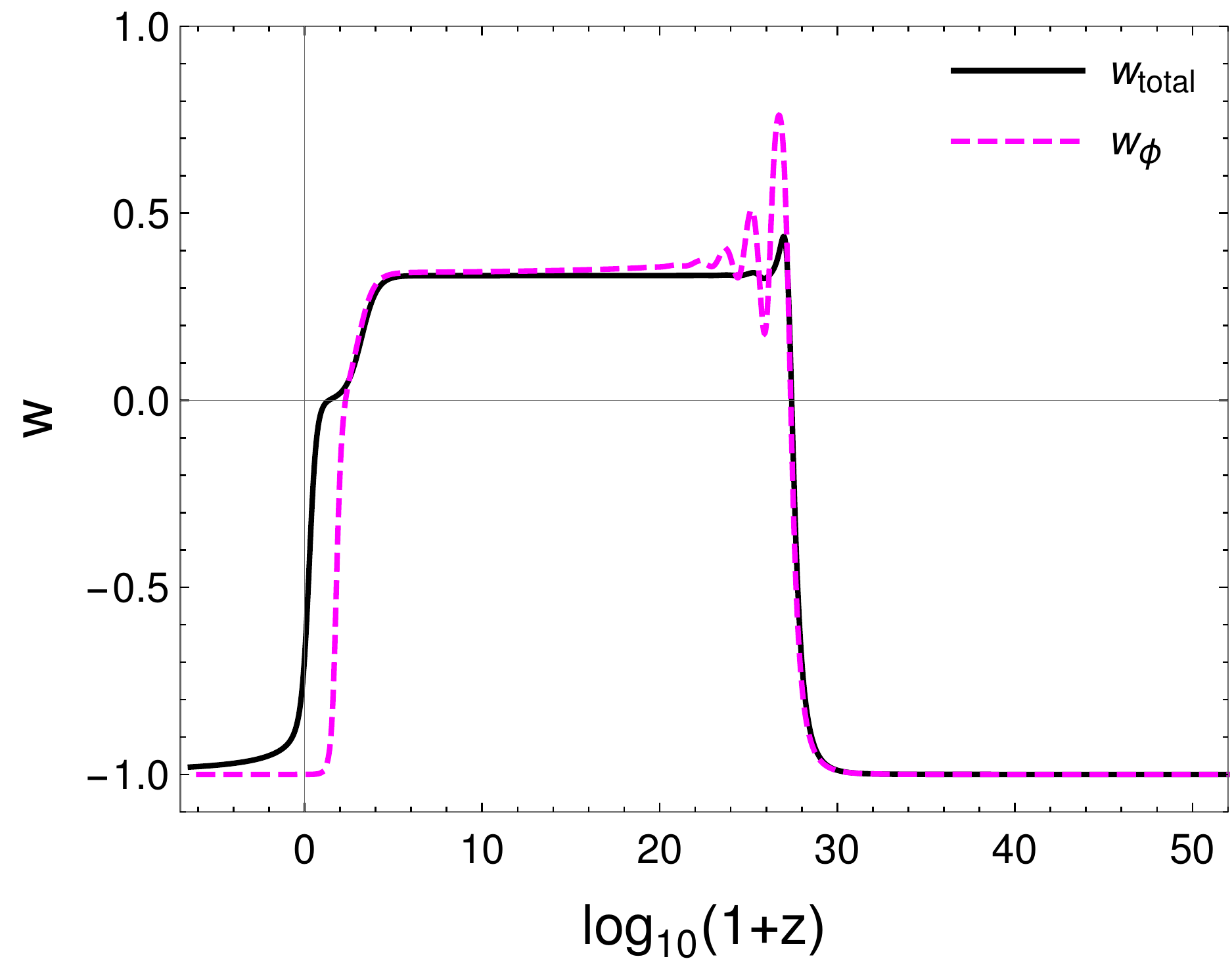}
\caption{The equations of state for the quintessence scalar field,
  $w_\phi$, and for the whole Universe, $w_{\rm total}$, as a function
  of the redshift, for the model $n=3$ and   parameters given in
  Tab.~\ref{tab1}. }
\label{figws}
\end{figure}
\end{center}
%%%%%%%%%%%%%%%%%%%%%%%%%%%%%%%%%%%%%%%%

In {}Fig.~\ref{figws}, we show the equation of state (EOS) for the
quintessence scalar field, $w_\phi$, and for the whole Universe,
$w_{\rm total}$. Again, we show the results only for the model with
$n=3$, with the other two models having very similar results. We note
from the results shown in {}Fig.~\ref{figws} that the Universe goes
from the inflationary regime to the radiation-dominated one 
 with a very short intermediate
kination period, in which $1/3 < w_{\rm total} \leq 1$.
This is quite different from typical inflation
quintessential models, in which one usually finds a rather long kination
period and the Universe
expands like stiff matter, $\rho \propto 1/a^6$. 
Here, however, because
of the different dissipative effects, first from the WI intrinsic
dissipation, Eq.~(\ref{cubic}), and the subsequent quintessence-dark 
matter interaction, Eq.~(\ref{Upsilonm}), their effects are
fundamental to first leading to a smooth transition from the
inflationary regime to the radiation-dominated one and then to
maintain the Universe in a radiation EOS $w_{\rm total}
= 1/3$, subsequently, except for a very short period after inflation
\footnote{The oscillating behavior in the EOS for the
  scalar field at the end of the inflationary 
phase and commencement of the radiation one (which also manifests with less intensity in $w_{\rm
    total}$), is a reminiscence of the kination tendency typical of
  the quintessential inflation models of the type studied here 
(this very small kination regime is also seen in {}Fig.~\ref{figenergies},
and discussed in there).
These
  features become less prominent for larger values of $n$ and they
  also tend to be damped away the larger is the dissipation ratio
  $Q_*$, which leads to a much smoother transition between the end of
  inflation to the radiation-dominated regime.}, as shown in
         {}Fig.~\ref{figws}, with $w_{\rm total}$ never reaching 1,
         as expected for a fully developed kination period.
This is quite an
         important result, since large kination periods can be
         potentially dangerous in many quintessence models. In a model
         like that given by Eq.~(\ref{pot}), in the absence of particle
         production mechanisms like the one of WI, one has to rely on
         gravitational particle production. The produced gravitational
         radiation can then be boosted by the kination period and it
         can potentially dominate by the time of big bang
         nucleosynthesis and can even challenge a successful
         nucleosynthesis~\cite{Ahmad:2017itq,Sahni:2001qp}.  It is
         quite satisfying to see that in the present model we do not
         have any such issues. In fact, we do not  have to rely at
         all on gravitational particle production. Particle production
         in the present case is all from quasiadiabatic motion of the
         scalar field $\phi$ along its potential. This mechanism of
         particle production, typical of the WI scenario (contrary to
         particle production from the quantum vacuum) is assured by
         the presence of the thermal bath, which was first
         discussed in connection to inflation by  Hosoya and Sakagama
         in Ref.~\cite{Hosoya:1983ke} and also by Morikawa and Sasaki
         in Ref.~\cite{Morikawa:1984dz}.  A proper quantum field
         theory interpretation of this particle production mechanism
         in a thermal bath was later given by Moss and Graham in
         Ref.~\cite{Graham:2008vu}, in particular leading to a
         dissipation coefficient of the form of Eq.~(\ref{cubic}).
{}For an additional remark concerning the results for the EOS
for the quintessence scalar field, $w_\phi$, and shown in {}Fig.~\ref{figws},
let us note that a time-varying  EOS for dark energy, of the form $w(a) = w_0 +(1-a) w_a$,
has been assumed by several works (see, e.g., Ref.~\cite{Aghanim:2018eyx}).
Given that the quintessence field in our model follows a quick freezing
solution after radiation domination and still in the matter dominated regime,
as can be noticed from {}Fig.~\ref{figws}, we have that $w_0 \simeq -1$
and $w_a\simeq 0$ and, then, the scalar field $\phi$ late-time behavior,
by the present epoch ($a_0=1$), is practically indistinguishable from that of a cosmological
constant.

%%%%%%%%%%%%%%FIGURE07%%%%%%%%%%%%%%%%%%%
\begin{center}
\begin{figure}[!htb]
\includegraphics[width=8.6cm]{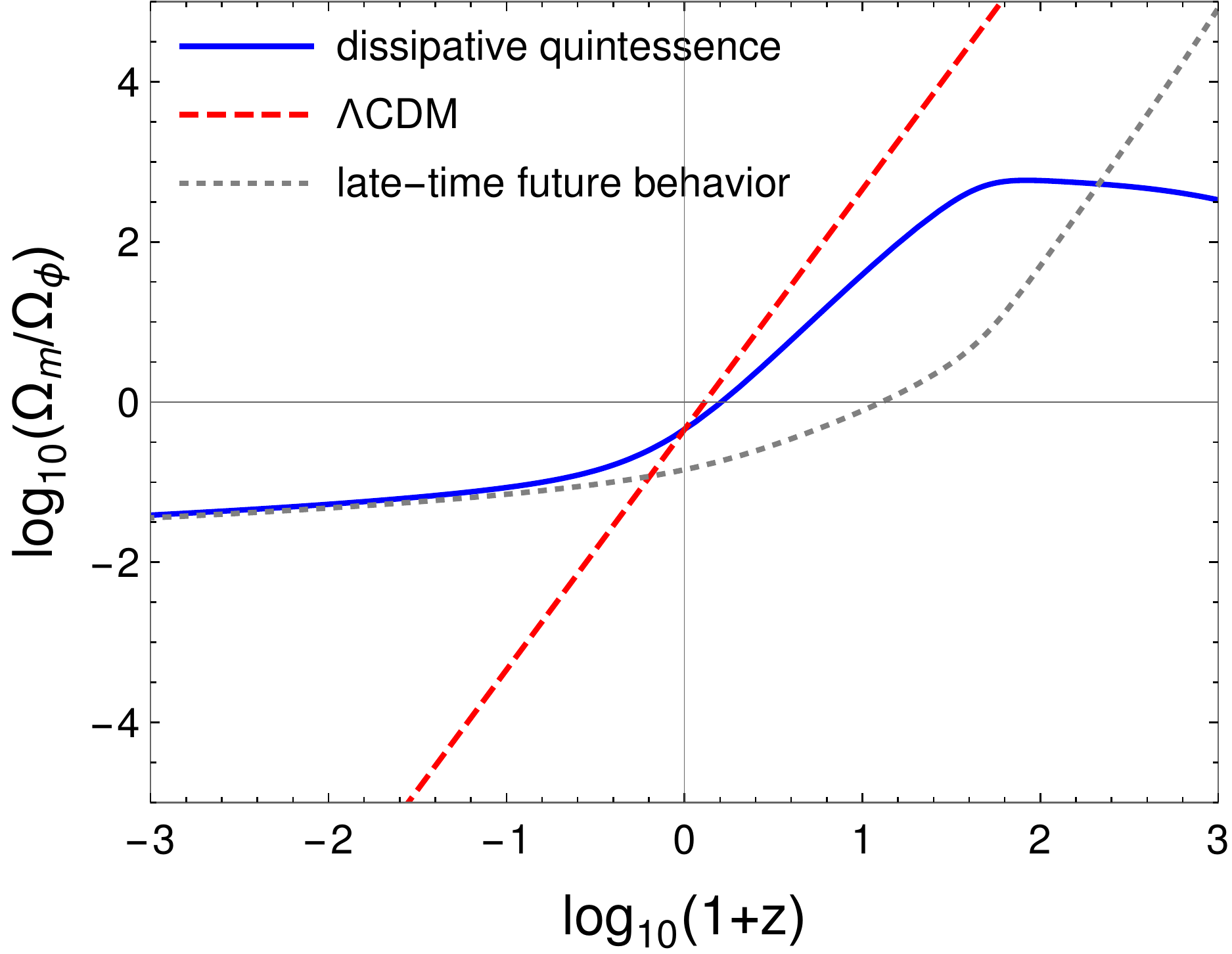}
\caption{The ratio of dark matter and quintessence energies densities,
  $\rho_{m}/\rho_\phi \equiv \Omega_m/\Omega_\phi$, as a function of
  the redshift, for the model $n=3$ and   parameters given in
  Table~\ref{tab1} (solid line) and the analogous quantity for the
  $\Lambda$CDM model (dashed line). The dotted line indicates the late-time
result given by Eq.~(\ref{latetime}).}
\label{figrhoratio}
\end{figure}
\end{center}
%%%%%%%%%%%%%%%%%%%%%%%%%%%%%%%%%%%%%%%%

{}Finally, in {}Fig.~\ref{figrhoratio}, we show the ratio of dark
matter and quintessence energies densities, $\rho_{m}/\rho_\phi \equiv
\Omega_m/\Omega_\phi$, as a function of the redshift and compare the
result with the one from the $\Lambda$CDM model. We notice that the
slope of the curve from the quintessence model $n=3$ (again the other
two cases shown in Table~\ref{tab1} give similar results) is smaller
than the one from the $\Lambda$CDM model. In particular, $
d(\rho_{m}/\rho_\phi)/dt|_{z_0=0} < H_0$, which has been seen also in
dark energy-dark matter interaction
models~\cite{delCampo:2008sr,Wang:2016lxa}.  The variation
$d(\rho_{m}/\rho_\phi)/dt$ in the present model being slower than that
of the $\Lambda$CDM model can alleviate the coincidence problem seen
in the latter~\cite{delCampo:2008sr,Wang:2016lxa}. 
We also note from {}Fig.~\ref{figrhoratio} that the scalar field does not
quite follow a scaling dynamics with the matter density in the matter dominated
regime, which is consistent with what is seen also in {}Fig.~\ref{figws}.
After matter-radiation equality the scalar field loses the scaling with the
dominant energy density. However, as we move toward the very distant future,
$z<0$, a late-time behavior relating $\rho_m$ with $\rho_\phi$ emerges. 
As we extend the dynamics toward the future, $z<0$, a slow-roll approximation for 
the background dynamics for $\phi$ and $\rho_m$ can be used. 
In this case, as the dissipation ratio $Q$ and
the radiation energy density $\rho_R$ are completely negligible,
we can approximate the Eqs.~(\ref{eqphinew}) and (\ref{eqrhom}), respectively, as
\begin{eqnarray}
&& \dot{\phi} \approx - \frac{V_{,\phi}}{\Upsilon_{\rho_m}} \approx - \frac{V_{,\phi}}{M^2} \rho_m^{1/4},
\label{dotphi}
\\ && \rho_m \approx \frac{\Upsilon_{\rho_m}}{3H} \dot \phi^2 \approx \frac{(V_{,\phi})^2}{3H M^2} \rho_m^{1/4},
\label{rhomslow}
\end{eqnarray}
where we have used in the above equations that at late times the dominant contribution
in $\Upsilon_{\rho_m}$ is the last term in Eq.~(\ref{Upsilonm}).
Thus, using that at late times $z<0$, $H^2 \sim \rho_\phi/(3 M_{\rm Pl}^2)$ and $\rho_\phi \sim
V(\phi)$, hence, from Eq.~(\ref{rhomslow}), we have that
\begin{equation}
\rho_m^{3/2} \approx  \frac{(V_{,\phi})^4 M_{\rm Pl}^2}{3 M^4 V(\phi)},
\label{rhomslow2}
\end{equation}
and, using also the expression for the potential Eq.~(\ref{pot}), we can finally obtain that
\begin{equation}
\frac{\rho_m}{\rho_\phi} \approx \left( \frac{\alpha \, n \, \phi^{n-1}}{M_{\rm Pl}^{n-1}} \right)^{8/3} 
\frac{V(\phi)}{(3 M^4 M_{\rm Pl}^2)^{2/3}}.
\label{latetime}
\end{equation}
{}From {}Fig.~\ref{figrhoratio} we see that the result given by Eq.~(\ref{latetime}), indicated
by the dotted line, agrees quite well
with the full numerical result when ${\rm log}_{10}(1+z) < -1$. 

%%%%%%%%%%%%%%%%%%%%%%%%%%%%%%%%%%%%%%%%%%%%%%%%%%%%%%%%%%%%%%
\section{Conclusions}
\label{sec5}

In this work we have proposed a dissipative quintessential inflation
model that is able to fully describe both the early inflationary
Universe and the present epoch. The early Universe dynamics is
realized in the warm inflation scenario. The late dynamics is
motivated by the recent dark energy-dark matter interaction models,
however, differently from what is adopted in those models, here we
motivate the interaction term by the dissipative terms typically found
in warm inflation.  By doing so, we have defined a new type of
interacting model connecting both early- and late-time dynamics. Both
the inflationary and the late-time periods have been described in
terms of a steep class of generalized exponential potentials. The role
of the different forms of dissipation terms adopted have been shown to
be fundamental to lead to a consistent picture. 

We have obtained values for the tensor-to-scalar ratio and the
spectral tilt of the primordial spectrum that are  consistent with the
ones obtained by the Planck legacy data. Likewise, we have an
appropriate  description of the late-time dynamics that avoids any
strong fine-tuning of parameters, as typically plagues many
quintessence models in general. Our results share many similarities
with other dark sector types of interaction models, including an
alleviation of the coincidence problem. 

Our model also displays some very welcome features.
{}For instance, the
WI dynamics allows for a smooth transition  from inflation to radiation
domination, while the scalar quintessence-dark matter interaction
avoids the
possible dangerous kination regime that could last for too long a period.
The model also allows for a proper
transition from dark matter to dark energy, including an appropriate
duration of the matter domination period and the transition to the
accelerated regime toward today's epoch.     

Interaction models, in general, have been studied also in the context of
dynamical systems (see, for example,
Refs.~\cite{Amendola:2006qi,Chimento:2007yt,Quartin:2008px} and for a
recent review, see Ref.~\cite{Bahamonde:2017ize}).  Since analytical
solutions can only be obtained in some very special and particular
cases, a  dynamical system analysis can allow one to obtain a qualitative
picture of the problem, including the stability, scaling solutions and
attractor points and also giving a qualitative view of the long-time
behavior of the cosmological dynamics. In the present work we have
adopted a more direct approach, by directly studying the background
dynamics numerically. It would, however, be important to study the
model  presented here also in a dynamical system context. Here, we
have performed an analysis for the predictions  from
perturbations only for the cosmological inflationary regime. Extending
such an analysis also for the late Universe dynamics (like computing the
redshift-space distortion $f\sigma_8$ and a comparison with the
analogous results from the $\Lambda$CDM model), including also a
statistical analysis, would certainly  help to better constrain the
parameters of the model. We expect to make these different
explorations of the model in the future.

%%%%%%%%%%%%%%%%%%%%%%%%%%%%%%%%%%%%%%%%%%%%%%%%%%%%%%
\section*{Acknowledgments}

G.B.F.L. is supported by a scholarship from Conselho
Nacional de Desenvolvimento Cient\'{\i}fico e Tecnol\'ogico (CNPq).
R.O.R. is partially supported by research grants from CNPq,
Grant No. 302545/2017-4, and Funda\c{c}\~ao Carlos Chagas Filho de
Amparo \`a Pesquisa do Estado do Rio de Janeiro (FAPERJ), Grant
No. E-26/202.892/2017.

%%%%%%%%%%%%%%%%%%%%%%%%%%%%%%%%%%%%%%%%%%%%%%%%%%%%%%%%%%%%%

\end{document}